\documentclass[twocolumn,twoside,slac_two]{revtex4}
\usepackage{graphicx}
\usepackage{fancyhdr}
\newcommand{\met}       {\mbox{$\not\!\!E_T$}}
\newcommand{\ttbar}     {\mbox{$t\bar{t}$}}
\newcommand{\sigmatt}   {\mbox{$\sigma_{t\bar{t}}$}}

\begin{document}

\title{Tevatron Top Results}

%

\author{C. Cl\'ement}
\affiliation{University of Stockholm, Fysikum 106 91 Stockholm , 
Sweden\footnote{Now at CERN, CH-1211 Gen\`eve 23, Switzerland,\\ cclement@cern.ch}}

\begin{abstract}
I present the latest results from the CDF and D\O\ collaborations
on top quark production (single top and top quark pair production) 
at the Tevatron $p\bar{p}$ collider at $\sqrt{s}$=1.96 TeV, measurements 
of the top quark decay properties such as the branching ratio 
${\cal B}(t\to Wb)$, the $W$ helicity in $t \to Wb$ decays, and 
measurements of fundamental parameters such as the top quark charge 
and mass.
\end{abstract}

\maketitle

\thispagestyle{fancy}


\section{Introduction}
The top quark was discovered~\cite{topdiscovery} by the CDF and D\O\ 
experiments~\cite{cdfupgrade,d0detector} in $p\bar{p}$ collisions at 
the Tevatron Run I at $\sqrt{s}=$1.8 TeV. The Tevatron Run II, which 
started in 2001, provides now collisions at $\sqrt{s}=$1.96 TeV 
with luminosities in the range $0.2-2 \cdot 10^{32}\rm{cm^{2} s^{-1}}$ 
and is expected to provide data-sets of several $fb^{-1}$ before the 
start of the CERN's Large Hadron Collider (LHC). The top quark is of 
particular interest because it is the heaviest fermion. Its mass 
is remarkably close to the electroweak scale and it 
has been speculated that the top quark could be related to the 
electroweak symmetry breaking mechanism~\cite{ewsb}. It is therefore 
important to measure its properties, not only to characterize it, 
but also to understand whether it is a mere part of the standard 
model (SM) or if it is the first sign of new physics beyond it. 
By virtue of its large mass, the top quark could also decay into 
exotic particles, e.g. a charged Higgs boson~\cite{charged_higgs}. 
The top quark has a lifetime of about $0.5 \cdot 10^{-24}$s,
much shorter than any other quark. In fact this lifetime is shorter
than the QCD hadronization time scale and the top quark decay is 
about as close as one can get to the decay of a free quark.
Understanding the top quark is also important because it will
be used as a calibration signal by the LHC experiments and might 
become the main background to many potential new physics signals, 
such as supersymmetry. The Run II of the Tevatron with the CDF 
and D\O\ experiments are now providing much larger dataset and allow for a 
series of precision measurements of the properties of the top quark.

\section{Top Quark Pair Production}
\begin{figure}[h]
\centering
\includegraphics[width=70mm]{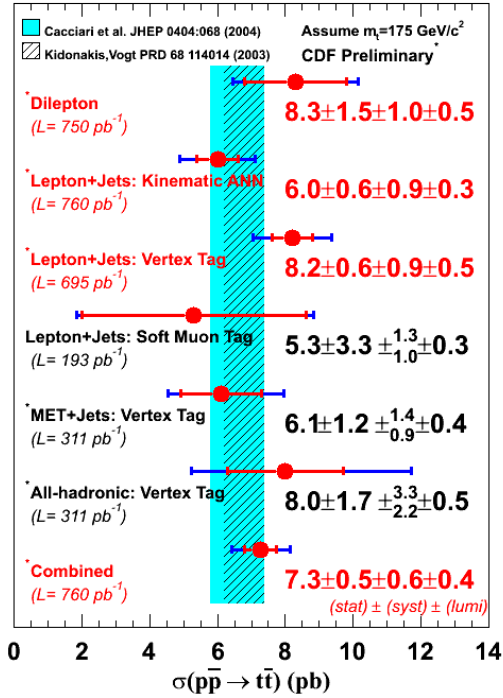}
\caption{Summary of the measured top pair production cross sections measured at CDF.} \label{fig:cdf_xs}
\end{figure}

\begin{figure}[h]
\centering
\includegraphics[width=70mm]{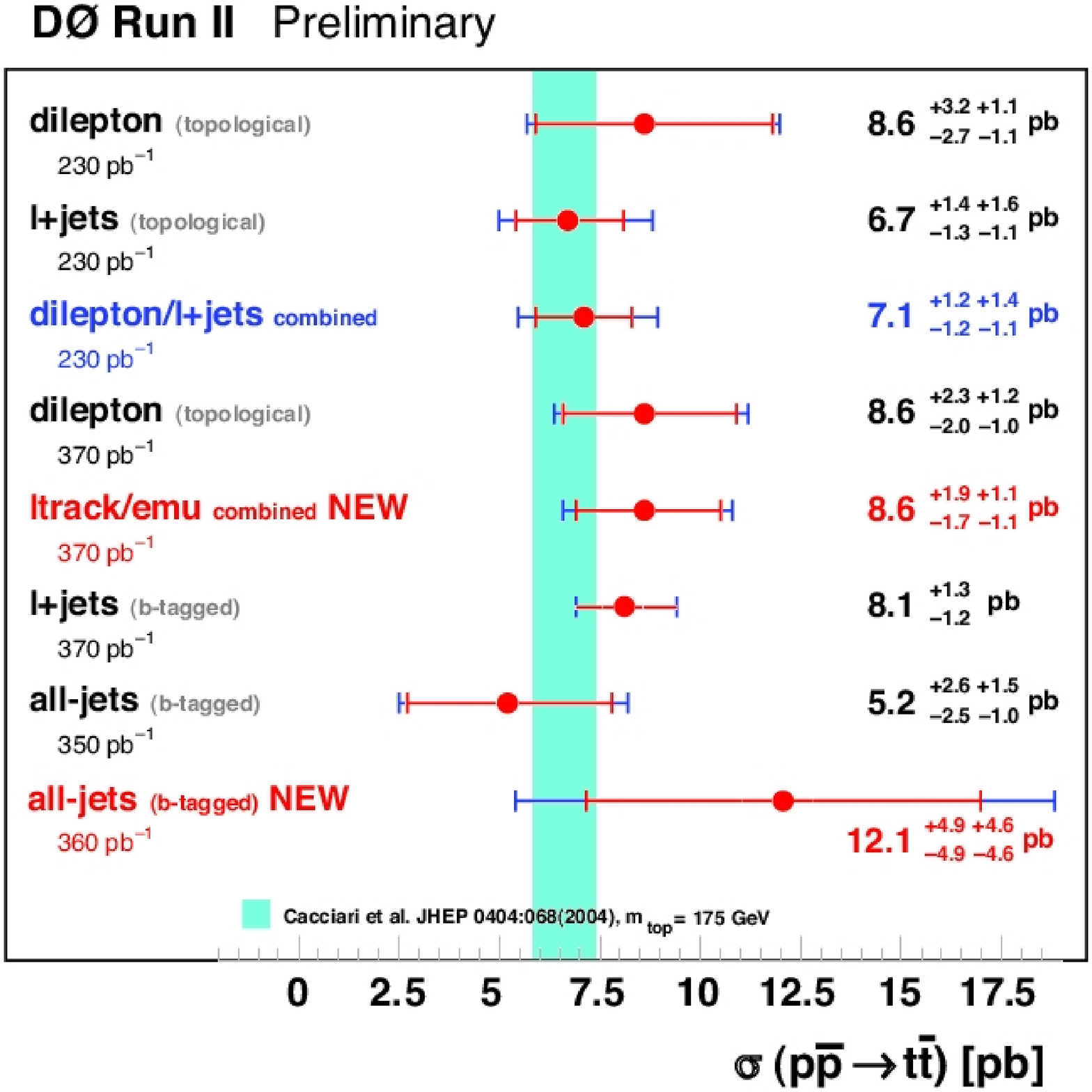}
\caption{Summary of the measured top pair production cross sections measured at D\O\ .} \label{fig:d0_xs}
\end{figure}

Top quarks can be produced in pairs (\ttbar) via the strong 
interaction, mainly via $q\bar{q}$ annihilation at the Tevatron energy.
Only pair-produced top quarks have been observed to date, though search 
for single top quark production is intense at CDF and D\O\, as detailed 
in next section. The theoretical cross section for $t\bar{t}$ production
is \sigmatt=6.78$\pm$1.2~pb~\cite{ttbar_xs}. Measuring
$\sigmatt$ is a test of pQCD at high $Q^2$. It should be 
sensitive to exotic top quark decays, since part of the 
top quark width could "disappear" into exotic channels, 
for which standard analyzes are not designed. Additional 
sources of $t\bar{t}$, beyond the SM could also be present,
leading to higher values of $\sigmatt$. CDF and D\O\ have 
measured $\sigmatt$ in a range of final states and using 
various techniques. Figures~\ref{fig:cdf_xs},~\ref{fig:d0_xs} 
summarize the measured $\sigmatt$ by the CDF and D\O\ 
collaborations. The most precise measurements are now 
limited by systematic uncertainties, from the luminosity 
measurement and the jet energy scale~\cite{jes_cdf,jes_d0}. 
The most competitive measurements are carried out in the 
so-called ``lepton-plus-jets'' channel, where one 
of the $W$ boson decay leptonically and one decays hadronically, 
giving rise to an average of four jets, one isolated high 
$p_T$ lepton and missing transverse energy ($\met$). The value 
of this channel comes from its relatively high branching 
fraction ($\sim$ 30\% of \ttbar\ final states), compared 
to the final state where both $W$'s decay leptonically (so-called
``dilepton'' channels). It can also be reasonably well separated 
from the dominant $W$+jets background, in comparison to the 
all-hadronic \ttbar\ channel, which has larger branching 
fraction but contains two hadronically decaying $W$'s and
not high $p_T$ lepton.
%
The dilepton final states offer the cleanest signature with two 
high $p_T$ leptons, but suffer from lower branching ratio 
($\sim$ 5\% of the \ttbar\ final states), making measurements
in this final state statistically limited. Interesting techniques 
have been developed to enhance the efficiency for the dilepton 
channels~\cite{ltrack_cdf,ltrack_d0} by identifying one of the 
leptons purely from its isolated track in the CDF or D\O\ 
central tracking system.

\section{Single Top Quark Production}
\begin{figure}[h]
\centering
\includegraphics[width=35mm]{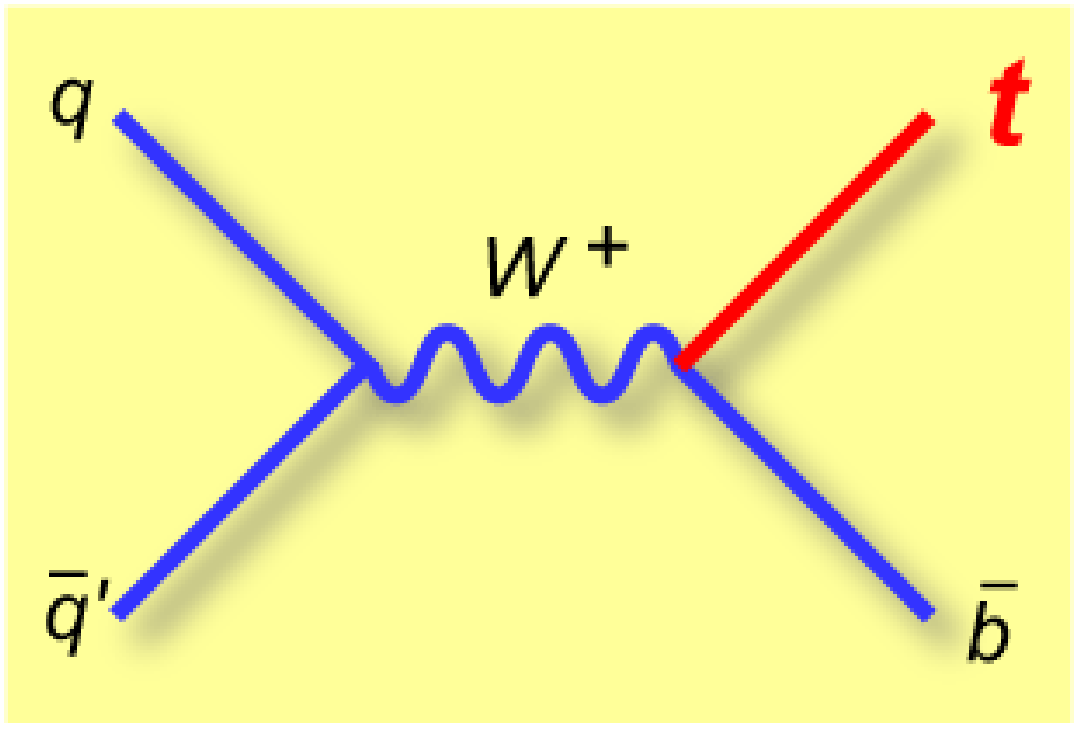}
\includegraphics[width=35mm]{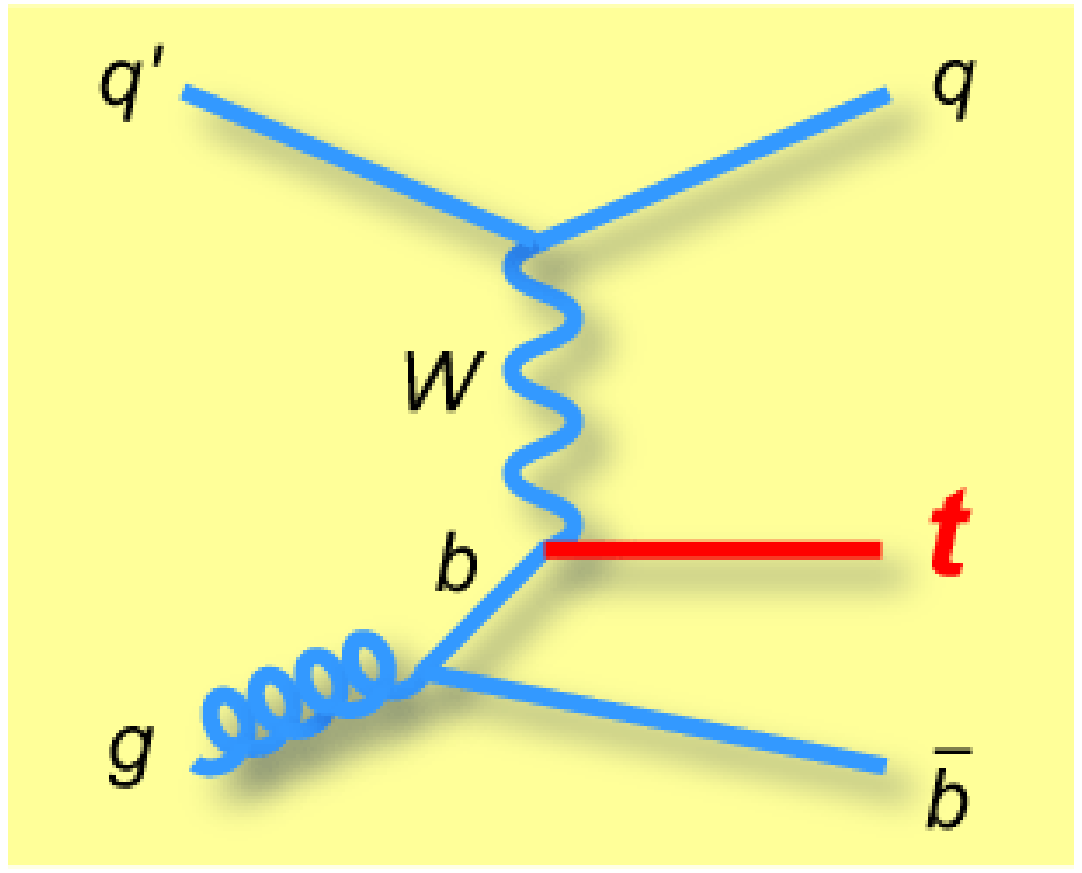}
\caption{Dominant processes for single top quark production: $s$-channel (left) 
and $t$-channel (right).} \label{fig:single_top}
\end{figure}

The production of a single top quark is possible 
through weak interaction, mainly via the processes 
pictured in Fig.~\ref{fig:single_top}. The cross section 
for these processes are $\sigma_t$=2.0~pb in the $t$-channel 
and $\sigma_s$=0.8~pb in the $s$-channel~\cite{singletop_xs}. 
The total single top production cross section is not significantly 
smaller than $t\bar{t}$ production cross section, but is 
significantly more challenging to isolate from backgrounds, 
due to fewer jets in the final states, and the presence of only 
one heavy object, instead of two, for $t\bar{t}$.

Single top production is important to search for, as it allows 
to test that indeed, as predicted by the SM, top quarks can be 
produced via electroweak interaction. An experimental
 determination of $\sigma_s$ and $\sigma_t$ would  provide a 
direct measurement of the CKM matrix element $|V_{tb}|$, since 
$\sigma_s$ and $\sigma_t$ are proportional to $|V_{tb}|^2$. 
Single top quark production could also be sensitive to new physics 
such as extra gauge bosons or additional quarks~\cite{single_top_np}.
Currently, in absence of single top signal, CDF and D\O\ place 
upper limits on $\sigma_s$ and $\sigma_t$, which are approaching 
the region of the cross sections predicted by the SM
for the $t$-channel. The latest measurements are summarized in 
Table~\ref{tab:single_top}. The single top signal is searched 
for in events where the $W$ boson decays leptonically, therefore 
giving rise to a hight $p_T$ isolated lepton and high $\met$. 
The $s$-channel has two $b$-quark jets, whereas the $t$-channel 
typically has one $b$-quark with a light-quark jet. The 
$\bar{b}$-quark in the $t$-channel is usually emitted in the 
forward direction with low $p_T$ and is often undetected. Therefore 
the two channels have different signatures and the search for 
the two channels are carried in general separately. The dominant 
backgrounds are \ttbar\ and $Wb\bar{b}$. The most common analysis 
method (referred as ``2D'' in Table~\ref{tab:single_top}) relies 
on discriminants, either likelihood-based (LH) or neural 
network-based (NN). For each channel ($s$ or $t$) one builds two 
discriminants {\cal D}$_1$ between signal and \ttbar\ and 
{\cal D}$_2$ between signal and $Wb\bar{b}$. Then the expected 
and observed 2D distributions of {\cal D}$_1$ versus {\cal D}$_2$ 
are used to derive the limit on $\sigma_s$ and $\sigma_t$.

\begin{table}[h]
\begin{center}
\caption{Upper limits at 95\% C.L. on single top production 
cross sections obtained by CDF~\cite{cdf_single_top_limit} and 
D\O\ ~\cite{d0_single_top_limit}.}
\begin{tabular}{llcc}
\hline 
\hline
                   & Analysis                   & Expected & Observed \\
\hline
CDF 695 pb$^{-1}$  & LH: $t+s$     & 5.9 pb & 3.3 pb \\
                   & 1D NN: $t+s$  & 5.7 pb & 3.4 pb \\
                   & 2D NN: $t$    & 4.2 pb & 3.1 pb \\
                   & 2D NN: $s$    & 3.7 pb & 3.2 pb \\
\hline
D\O\ 370 pb$^{-1}$ & 2D LH: $s$    & 3.5 pb & 5.0 pb \\
                   & 2D LH: $t$    & 4.3 pb & 4.4 pb \\
\hline
D\O\ 240 pb$^{-1}$ & 2D NN: $s$    & 4.5 pb & 6.4 pb \\
                   & 2D NN: $t$    & 5.8 pb & 5.0 pb \\
\hline
\hline
\end{tabular}
\label{tab:single_top}
\end{center}
\end{table}

\section{Top Quark Branching Fraction to $\mathbf{\sl Wb}$}
The top quark decays predominantly via $t\to Wq$, where $q=d,s,b$. 
Flavor changing neutral current decays of the type $t\to Vq$, where 
$V=g,Z,\gamma$ and $q=u,c$ are or the order of $10^{-10}$ or smaller 
within the SM~\cite{fcnctop}. Assuming exactly three generations of 
quarks, the $3\times3$ quark mixing matrix is unitary, and using the 
experimentally measured values of $V_{ub}$ and $V_{cb}$~\cite{vub_vcb}, 
one obtains $0.9990<V_{tb}<0.9992$. This also leads to 
${\cal B}(t\to Wb)\simeq$ 100\%. A deviation from the SM prediction 
could arise in the presence of a $4^{th}$ quark generation, or 
contamination of the sample by other processes than \ttbar\ .
CDF and D\O\ have both measured ${\cal B}(t\to Wb)$ by looking at 
the fraction of $t\bar{t}$ events with 0, 1 or 2 $b$-quark jets. 
Both analyzes assume that $t\to Xq$, where $X\neq W$, is negligible. 
In events with exactly zero $b$-quark jets, the S/B is low 
($\sim$ 1/10), but it is improved with discriminant techniques 
(artificial neural network for CDF and a likelihood discriminant 
for D\O\ ) that uses the kinematic properties of the $t\bar{t}$ 
events. CDF combines both lepton-plus-jets and dilepton final states, 
D\O\ uses only lepton-plus-jets final states but fits both the 
$t\bar{t}$ content and ${\cal B}(t\to Wb)$ simultaneously. Both 
experiments derive lower limits on $|V_{tb}|$ assuming 
$|V_{tb}|=\sqrt{{\cal B}(t\to Wb)}$. The results~\cite{cdf_btwb,d0_btwb} are 
summarized in Table~\ref{tab:btwb}. D\O\ has also shown that 
$\sigmatt$ and ${\cal B}(t\to Wb)$ can be measured simultaneously,
which helps to reduce the systematic uncertainties on the top pair 
production cross section and ${\cal B}(t\to Wb)$~\cite{btwb_2D}.

\begin{table}[h]
\begin{center}
\caption{Summary of CDF and D\O\ measurements of ${\cal B}(t \to Wb)$ and 95\% 
C.L. lower limits on $|V_{tb}|$.}
\begin{tabular}{lcc}
\hline 
\hline
                   & ${\cal B}(t\to \sl Wb)$ & $|V_{tb}|$ lower limit \\
\hline
CDF  160 pb$^{-1}$ & 1.12$^{+0.27}_{-0.23}$ & $|V_{tb}|>$0.78 \\
D\O\ 240 pb$^{-1}$ & 1.03$^{+0.19}_{-0.17}$ & $|V_{tb}|>$0.78 \\
\hline
\hline
\end{tabular}
\label{tab:btwb}
\end{center}
\end{table}

\section{$\mathbf{W}$ Helicity in Top Decays}
\begin{figure}[h]
\centering
\includegraphics[width=70mm]{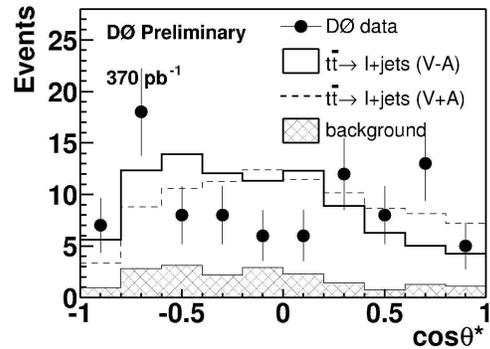}
\caption{Expected and observed distributions of $\cos{\theta^{*}}$ for
370 pb$^{-1}$ of D\O\ lepton-plus-jets data.}\label{fig:d0_helicity}
\end{figure}

In the SM the top quark decays via the V-A electroweak current 
according to
\begin{equation}
  \frac{-ig}{2\sqrt{2}} \bar{b} \gamma^{\mu} (1-\gamma^5) |V_{tb}| t W_{\mu}
\end{equation}
where the $\frac{(1-\gamma^5)}{2}$ operator has the effect of reducing the 
$b$-quark to its left-handed component. In practice, since 
the mass of the $b$-quark is negligible compared to the top 
quark mass, the top quark decays predominantly to left-handed 
$b$-quarks and by angular momentum conservation to left-handed 
$W$-bosons. In the SM the fraction of left-handed 
$W$'s ($f^-$) is expected to be $\sim$ 70\%, and $\sim$ 30\%
for $f^0$, the fraction of longitudinally polarized 
$W$s. The SM predicts that the fraction $f^{+}$ of right-handed 
$W$'s is $3.6 \cdot 10^{-4}$. A deviation from these predictions 
could indicate non-SM physics such as large CP-violation in 
top quark decays~\cite{helicity_motivation}. The polarization 
of the $W$ determines the angular distribution of emission of 
the lepton in the $W\to \ell \nu$ decay. Therefore one approach
to measure $f^{\pm}$ and $f^0$ is to measure the distribution
of $\cos{\theta^*}$, where $\theta^*$ is defined as the angle 
of emission of the lepton in the $W$ rest frame, with respect 
to the top quark line of flight. D\O\ uses the observed 
distribution of $\cos{\theta^*}$ in 370 pb$^{-1}$ of 
lepton-plus-jets data (Fig.~\ref{fig:d0_helicity}) to fit 
the fraction $f^{+}$~\cite{d0_helicity}. The $W$ polarization 
affects the $p_T$ of 
the lepton, since for a left-handed $W$, $\cos{\theta^*}$ is closer 
to minus one, the lepton is emitted in general more anti-parallel
to the $W$ boost direction and therefore has a lower average 
momentum in the lab-frame than for a lepton coming from a
right-handed $W$. The CDF collaboration combines both the
observed $\cos{\theta^*}$ and lepton $p_T$ 
distributions~\cite{cdf_helicity}, using lepton-plus-jets 
and dilepton events to derive the an upper limit on $f^{+}$ 
and a measurement of $f^{0}$.
Results are summarized in Tab.~\ref{tab:whelicity}.

\begin{table}[h]
\begin{center}
\caption{CDF and D\O\ measurements of $W$ helicity in $t \to Wb$ decays.}
\begin{tabular}{lc}
\hline 
\hline
                   & Result \\
\hline
CDF  200 pb$^{-1}$ & $f^{+}<0.27$ at 95\% C.L.  \\
                   & $f^0=0.74^{+0.22}_{-0.34}$ \\
\hline
D\O\ 370 pb$^{-1}$ & $f^{+}=0.08\pm0.08\rm~(stat) \pm 0.05~(syst)$ \\
\hline
\hline
\end{tabular}
\label{tab:whelicity}
\end{center}
\end{table}

\section{Top Quark Charge}
It is widely believed that the heavy particle discovered by 
the CDF and D\O\ collaborations is the long-sought top quark. 
Still, it is possible to interpret the discovered particle 
as either a charge $2e/3$ or $-4e/3$ quark. In the published 
top quark analyzes of the CDF and D\O\ 
collaborations~\cite{top_review}, the correlations of the 
$b$-quarks and the $W$-bosons in the reaction  
$p \bar{p} \to t \bar{t}\to W^+W^- b \bar{b}$ are not uniquely 
determined. As a result, there is a twofold ambiguity in the 
pairing of $W$ bosons and $b$-quarks, and, consequently, in the 
electric charge assignment of the ``top quark''. In addition 
to the SM assignment, $t \to W^+ b$, $''t'' \to W^-b$ is 
also conceivable, in which case the top quark would
actually be an exotic quark with charge $q=-4e/3$. It is
possible to make electroweak fits of $Z\to \ell^+ \ell^-$ and $Z\to
b \bar{b}$ data assuming a top quark of mass $m_{t}=270$ GeV and
that the right-handed $b$-quark mixes with the isospin +1/2
component of an exotic doublet of charge $-1e$/3 and $-4 e$/3
quarks, $(Q1~,Q4)_{R}$~\cite{exotic_top_paper}. In this scenario,
the $-4 e$/3 charge quark is the particle discovered at the
Tevatron, and the top quark, with mass of 270 GeV, would have so far
escaped detection.

\begin{figure}[h]
\centering
\includegraphics[width=70mm]{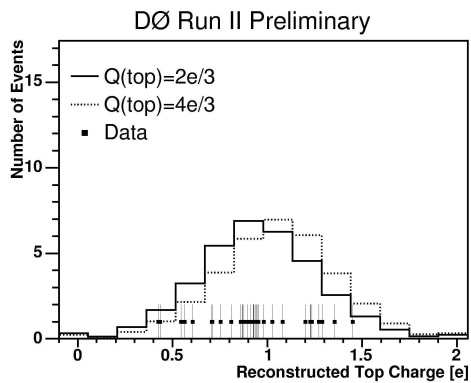}
\caption{Expected top quark charge distributions in D\O\ for 
the SM and exotic scenarios, together with the 
observed data point.}\label{fig:top_charge}
\end{figure}

The D\O\ collaboration has reported~\cite{top_charge} the 
first experimental discrimination between the $2e/4$ and 
$4e/3$ scenarios. D\O\ uses 370 pb$^{-1}$ of data, in the 
lepton-plus-jets channel with $b$-tagging 
techniques exploiting the long lifetime of $B$-hadrons. 
A very pure \ttbar\ sample is selected by requiring 
events with at least two $b$-tagged jets. 
A kinematic fit is performed on the events in order 
to fully reconstruct the \ttbar\ event. The $b$- and 
$\bar{b}$-jets are separated using a jet charge 
algorithm that combines the $p_T$ and charge of tracks 
inside the cone of the $b$-tagged jet. The expected 
distribution of jet charge for $b$- and $\bar{b}$-jets 
is largely derived from independent collider data-samples. 
Figure~\ref{fig:top_charge} shows that D\O\ finds the 
data in good agreement with the SM
expectation, and excludes the $4e/3$ scenario at the 
94\% C.L.

\section{Top Quark Mass}
The top quark mass $M_{top}$ is an important parameter of the 
SM, since it allows to predict the Higgs mass using precise 
measurements of the electroweak parameters. At Tevatron, the 
golden channel to measure $M_{top}$ is the lepton-plus-jets 
channel thanks to large statistics and significantly better 
S/B than the all hadronic channel. The CDF and D\O\ experiments 
treat separately \ttbar\ candidate events depending on the number 
of $b$-tagged jets, to fully exploit specific features of
each type of events. For instance double tagged events are 
fewer, but the S/B is large and the two $b$-tagged jets allow 
to significantly decrease the number of possible permutations
to reconstruct the \ttbar\ events. On the other hand, events with one 
or even zero $b$-tagged jets are less pure, but have larger 
statistics. It is optimal to exploit the specific features
of $b$-tag multiplicity bin independently and combine them. 
A relatively recent development in top quark mass measurements 
is the treatment 
of the jet energy scale. The uncertainty 
on $M_{top}$ in the lepton-plus-jets channel is dominated by 
systematic uncertainties. The dominant source is the jet 
energy scale. Each lepton-plus-jets event contains one 
$W\to q\bar{q}'$ decay, giving rise to two hadronic jets 
whose invariant mass must be consistent with the known $W$ 
mass. This can be used to simultaneously constrain the jet 
energy scale as one fits $M_{top}$ to the distribution of 
observed $M_{top}$. By carrying out this 2D fit, so
called ``in situ'' jet energy calibration, the systematic error
from jet energy scale can be reduced significantly, most of it
becoming of statistical nature, and therefore scaling as the
square root of the integrated luminosity. A residual jet energy
scale systematic uncertainty remains due to {\it i)} 
$b$-jet energy scale, known to be different at a couple of percents
level from the light-jet energy scale, and {\it ii)} the 
extrapolation of the jet energy scale from the $W$ mass 
to higher masses.

\begin{figure}[h]
\centering
\includegraphics[width=70mm]{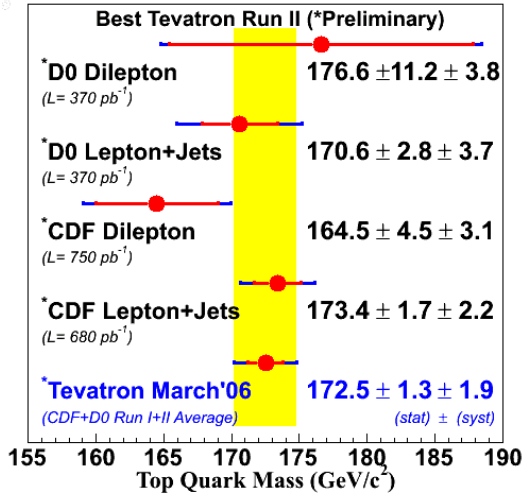}
\caption{Summary of measured top quark mass by CDF and D\O\ .}\label{fig:top_mass_summary}
\end{figure}

The currently most performant results from CDF and D\O\ are 
summarized in Table~\ref{tab:top_mass}. The CDF 
measurement~\cite{cdf_top_mass} uses the so-called template 
method~\cite{template_method}. The D\O\ measurement~\cite{d0_top_mass} 
uses the matrix element method described in Ref.~\cite{me_method}.
The top quark mass has also been measured also in dilepton and all 
hadronic channels~\cite{cdf_dilepton_mass,d0_dilepton_mass}.
The main measurements are summarized in Figure~\ref{fig:top_mass_summary}. 
These results are in good agreement with the electroweak precision 
measurements from LEP1 and SLD~\cite{top_mass_ew} which yields 
$M_{top}=172.6^{+13.2}_{-10.2}$ GeV.

\begin{table}[h]
\begin{center}
\caption{Current best Tevatron $M_{top}$ measurements in the lepton-plus-jets channel. For the analyzes using the in-situ jet energy scale calibration, the statistical part of the jet energy scale systematic uncertainty (JES) is grouped with the statistical uncertainty (stat).}
\begin{tabular}{ll}
\hline 
\hline
                    & Measured $M_{top}$ (GeV) \\
\hline
CDF 680 pb$^{-1}$   & $173.4 \pm 2.5 \rm~(stat+JES) \pm 1.3 ~(syst)$ \\
D\O\ 370 pb$^{-1}$  & $170.6 ^{+4.0}_{-4.7} \rm~(stat+JES) \pm 1.4 ~(syst)$ \\
\hline
Combined Run I + II & $172.5 \pm 1.3 \rm~(stat) \pm 1.9 ~(syst)$ \\
\hline
\hline
\end{tabular}
\label{tab:top_mass}
\end{center}
\end{table}

\begin{acknowledgments}
I would like to sincerely thank the Fermilab Beam Division staff 
for making possible the study of top quark, and my fellow colleagues 
from the CDF and D\O\ collaborations. I would like to acknowledge 
funding from The Swedish Research Council (diarienr 2004-3122) 
who made part of this work possible.
 
\end{acknowledgments}

\bigskip 

\end{document}